\documentclass{article}

 \usepackage[preprint]{neurips_2026}

\usepackage[utf8]{inputenc} %
\usepackage[T1]{fontenc}    %
\usepackage{hyperref}       %
\usepackage{url}            %
\usepackage{booktabs}       %
\usepackage{amsfonts}       %
\usepackage{nicefrac}       %
\usepackage{microtype}      %
\usepackage{xcolor}         %

\usepackage{tabularx}
\usepackage{dirtree}
\usepackage{listings}
\usepackage{graphicx}
\usepackage{multirow}
\usepackage[most, listings]{tcolorbox}

\usepackage{amsmath}
\usepackage{amssymb}        %
\usepackage{tikz}
\usetikzlibrary{arrows.meta, calc}
\usepackage{wrapfig} %

\usepackage{color}
\usepackage{enumitem}

\newtcolorbox{promptbox}[1]{
    colback=gray!5,     %
    colframe=gray!75,   %
    fonttitle=\bfseries,
    title=#1,           %
    arc=2pt,            %
    outer arc=2pt,
    boxrule=0.5pt,
    left=5pt, right=5pt, top=5pt, bottom=5pt
}

\newtcblisting{markdownprompt}[1]{
    listing only,          %
    colback=gray!5,
    colframe=gray!75,
    fonttitle=\bfseries\sffamily,
    title=#1,
    arc=2pt,
    outer arc=2pt,
    boxrule=0.5pt,
    left=5pt, right=5pt, top=5pt, bottom=5pt,
    breakable,
    listing options={
        basicstyle=\small\ttfamily,
        breaklines=true,
        columns=fullflexible,
        keepspaces=true,
        literate={—}{{---}}1
                 {–}{{--}}1
                 {↔}{{$\leftrightarrow$}}1
                 {×}{{$\times$}}1
                 {→}{{$\rightarrow$}}1
                 {├}{{\textsf{|--}}}1 
                 {─}{{\textsf{--}}}1 
                 {└}{{\textsf{\char124}--}}1 
                 {│}{{\textsf{|}}}1 
                 {←}{{$\leftarrow$}}1
                 {~}{{\ }}1, %
    }
}

\definecolor{mGreen}{rgb}{0,0.6,0}
\definecolor{mGray}{rgb}{0.5,0.5,0.5}
\definecolor{mPurple}{rgb}{0.58,0,0.82}
\definecolor{backgroundColour}{rgb}{0.95,0.95,0.92}

\lstdefinestyle{CStyle}{
    backgroundcolor=\color{backgroundColour},   
    commentstyle=\color{mGreen},
    keywordstyle=\color{magenta},
    numberstyle=\tiny\color{mGray},
    stringstyle=\color{mPurple},
    basicstyle=\ttfamily\footnotesize,
    breakatwhitespace=false,         
    breaklines=true,                 
    captionpos=b,                    
    keepspaces=true,                 
    numbers=left,                    
    numbersep=5pt,                  
    showspaces=false,                
    showstringspaces=false,
    showtabs=false,                  
    tabsize=4,
    language=C
}

\usepackage{cleveref}
\usepackage{subcaption}
\usepackage[table]{xcolor}

\title{HSCO-Bench: An Agent-Driven End-to-End Hardware-Software Co-design Benchmark for Systems-on-Chip}

\author{%
  Pei-Huan Tsai \\
  Columbia University\\
  \texttt{tph@cs.columbia.edu} \\
  \And
  Kuan-Lin Chiu \\
  Columbia University\\
  \texttt{chiu@cs.columbia.edu} \\
  \And
  William Baisi \\
  Columbia University \\
  \texttt{wb2426@columbia.edu} \\
  \AND
  Pin-Yu Chen \\
  IBM Research \\
  \texttt{pin-yu.chen@ibm.com} \\
  \And
  Luca P. Carloni \\
  Columbia University \\
  \texttt{luca@cs.columbia.edu} \\
}

\newcommand{\hsco}{{\text{HSCO-Bench}}}

\begin{document}

\maketitle

\begin{abstract}
Large language models (LLMs) are increasingly adopted for software and hardware design, yet these domains are still evaluated separately.
Software benchmarks typically assume fixed hardware targets, while hardware benchmarks focus on component-level optimization without considering the full hardware-software stack.
Consequently, no existing benchmark evaluates whether an LLM agent can perform end-to-end, system-level hardware-software co-design.
Such a process requires:
1) \emph{analyzing} applications to identify kernels requiring acceleration,
2) \emph{designing} and \emph{integrating} heterogeneous accelerators into a System-on-Chip (SoC) under resource constraints, and
3) \emph{mapping} kernels onto the generated accelerators.
We present \hsco{}, an end-to-end hardware-software co-design benchmark for accelerator-rich heterogeneous SoC generation.
Built upon an open-source SoC platform with a curated repository structure, \hsco{} evaluates the ability of LLMs to jointly optimize software and hardware stacks, producing SoC prototypes deployed on the AMD Virtex-7 FPGA VC707 Evaluation Kit.
Experimental results show that end-to-end integration remains challenging for current models, with widespread failures.
Among the five frontier models evaluated, only two of them could successfully generate valid SoC prototypes. Yet, even in these successful instances, the generated designs are far from optimal. While we observe a promising peak speedup of $16.22\times$, the maximum additional resource utilization reaches only $23.67\%$. This highlights that while state-of-the-art models demonstrate an emerging capability for hardware acceleration, they still heavily underutilize the available hardware capacity, leaving substantial room for future optimization.
To the best of our knowledge, \hsco{} is the first benchmark targeting this complete co-design flow, enabling LLMs to jointly reason about and modify both the software and hardware stacks of heterogeneous SoCs.

\end{abstract}

\section{Introduction}
\label{section:intro}

\begin{figure}[t]
\begin{center}
\includegraphics[width=\columnwidth]{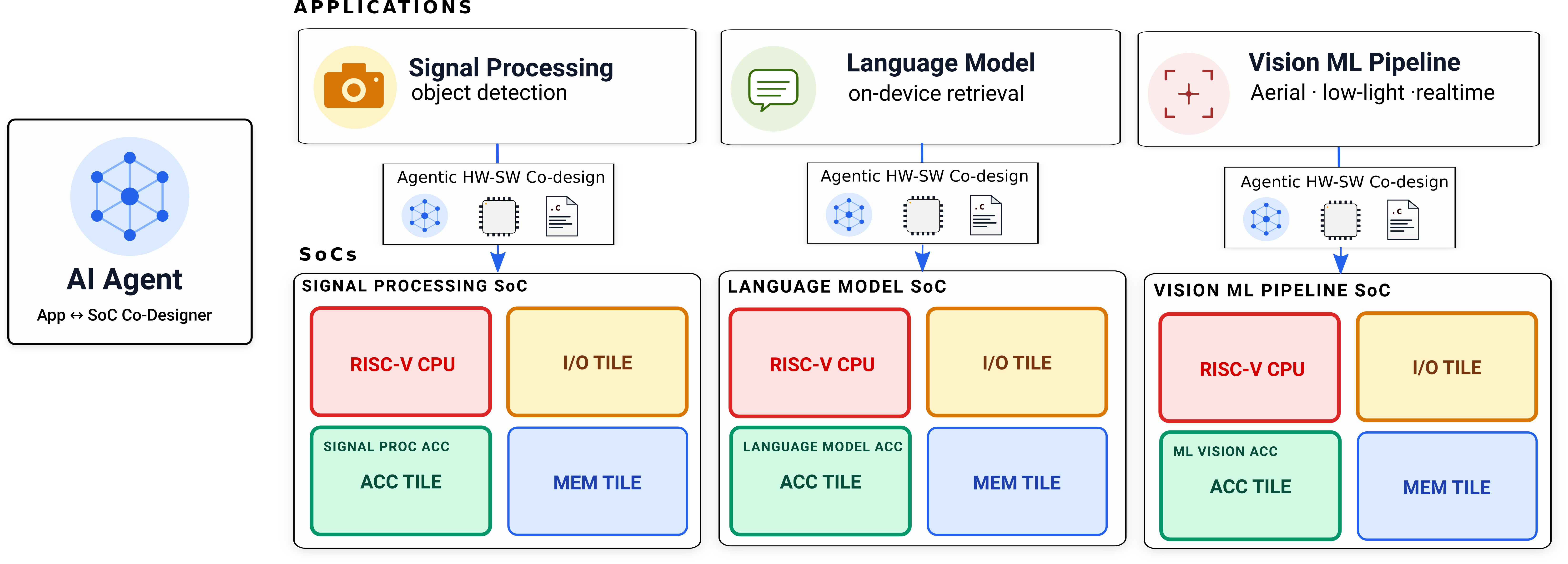}
\caption{Agent-powered autonomous SoC generation and hardware-software co-design}
\label{fig:banner}
\end{center}
\vspace{-15pt}
\end{figure}

Large language models (LLMs) are rapidly transforming software and hardware engineering.
Numerous benchmarks evaluate LLM capabilities across design, optimization, and verification tasks.
In software, benchmarks are proposed to test if LLM agents can resolve real-world GitHub issues~\citep{swebench}, write high-performance GPU kernels~\citep{kernelbench, tritonbench}, and generate formally verifiable code~\citep{verina_2025}.
In hardware, benchmarks are designed to evaluate whether LLMs can generate Verilog modules from natural language specifications~\citep{rtllm, verieval_v2}, design under resource constraints~\citep{resbench}, solve chip-level design problems~\citep{sldb, cvdp, chipbench}, and synthesize accelerators through high-level synthesis (HLS) from standalone kernel descriptions~\citep{hlsyn, hls_eval, hls_bench1}.

However, these two domains are typically evaluated in isolation.
Software benchmarks typically assume fixed hardware targets~\citep{kernelbench, tritonbench}, excluding hardware design from the optimization space.
Conversely, hardware benchmarks assume predefined computation targets, focusing on hardware optimization~\citep{rtllm,verieval_v2} without extending the design space to the application layer. In practice, the most consequential design decisions span this boundary, particularly within the development of heterogeneous multicore systems-on-chip (SoCs)~\citep{soc_shao, openpiton, chipyard, esp}. These architectures, which contain multiple specialized accelerators tailored for diverse applications to optimize execution performance, serve as the backbone of modern computing systems.
In such systems, software and hardware can be considered as a unified design space, requiring a strategic balance between specialized hardware and general-purpose computing components to fulfill computational tasks. Yet, there is no benchmark that encompasses:
1) end-to-end hardware-software co-design tasks on SoC platforms, enabling an LLM agent to guide its accelerator design and \emph{create its own specialized SoC} based on the software needs; and
2) system-level resource allocation and performance optimization tasks that require LLMs to \emph{jointly reason} about software workloads and hardware resources, a process that demands substantial human effort driven by experience, profiling, and iterative prototyping.

To address this gap, we introduce \hsco{}, the first benchmark for end-to-end hardware-software co-design and accelerator-rich SoC generation. Built upon the ESP open-source platform~\citep{esp}, \hsco{} challenges LLM agents to autonomously navigate the full design pipeline, enabling automated SoC generation (\figurename~\ref{fig:banner}).

Our contributions are as follows:
\begin{enumerate}
    \item \textbf{The First End-to-End Hardware-Software Co-Design Benchmark.} We formalize a comprehensive pipeline requiring an LLM agent to autonomously: \textbf{(i)} identify software kernels for acceleration; \textbf{(ii)} design and generate custom hardware via HLS; \textbf{(iii)} make system-level architectural decisions for SoC integration; and \textbf{(iv)} produce an FPGA-verified SoC. Unlike prior isolated benchmarks, \hsco{} uniquely evaluates an agent's capability to reason about cross-domain trade-offs and full-stack integration.
    \item \textbf{Curated Application Suite and Push-Button Evaluation Flow.} We construct a diverse suite of 10 applications spanning machine learning and signal processing. Our framework rigorously evaluates generated designs along two orthogonal axes: \textit{performance speedup} (measured in CPU cycles) and \textit{hardware overhead} (FPGA resource utilization, including logic cells, block RAMs (BRAMs), and digital signal processing slices (DSPs)).
    \item \textbf{Empirical Insights and Bottleneck Analysis.} Evaluating 5 frontier models driven by \texttt{mini-swe-agent}~\citep{swe_agent}, we demonstrate that LLMs can produce functionally correct, FPGA-verified SoCs. Notably, Opus 4.6 achieves a 70\% average end-to-end success rate and a geometric-mean speedup of $2.32\times$. Furthermore, our granular failure taxonomy identifies accelerator design and SoC integration as major bottlenecks, and reveals that even successful agents frequently underutilize available hardware resources, highlighting critical optimization gaps for future research.
\end{enumerate}

Our benchmark is available at~\url{https://github.com/B07901087/hsco_bench}

\section{Background and Related Work}
\label{section:prelim}

\subsection{Background}
\label{sec:prelim:hls}

\textbf{High-Level Synthesis (HLS).}
Hand-written RTL (e.g., Verilog/VHDL) requires deep micro-architectural expertise and long development cycles. High-level synthesis (HLS) addresses this gap by compiling untimed behavioral descriptions (typically C/C++ or SystemC) into synthesizable RTL, automatically inferring pipelining, loop unrolling, and memory interfaces~\citep{coussy_HLS_book}. From an LLM's perspective, HLS offers two key advantages. First, C-like syntax aligns with the vast software corpora LLMs are already fluent in, significantly lowering the barrier to hardware generation. Second, HLS makes the hardware/software boundary \emph{negotiable}: agents can flexibly move compute kernels in or out of hardware with minimal source-level changes.

\textbf{Heterogeneous SoC Platforms.}
Modern computing increasingly relies on heterogeneous SoCs integrating general-purpose cores with domain-specific accelerators due to the end of Dennard scaling and the slowdown of Moore's law~\citep{dally_dsa, horowitz_isscc}. While several open-source SoC platforms exist (e.g., OpenPiton~\citep{openpiton}, Chipyard~\citep{chipyard}), \hsco{} is built upon ESP~\citep{esp}. We select ESP because it combines two properties critical for an LLM-driven benchmark: (1) native support for HLS flows, which simplifies hardware design for agents; and (2)~a tile-based SoC architecture interconnected by a Network-on-Chip (NoC), where standardized \emph{sockets} abstract away low-level system integration challenges. This allows LLM agents to focus on high-level accelerator design and system-wide orchestration.

The combination of an extensible SoC platform and high-level hardware abstractions provides the \emph{automation} and \emph{flexibility} needed to evaluate full-stack reasoning in modern LLM agents.
While ESP serves as a hardware-validated testbed, the capabilities evaluated in \hsco{}, including kernel selection, accelerator design, and system-level integration, generalize to modern SoC architectures.

\subsection{Related Work}

\textbf{LLM Benchmarks for Software Engineering and Kernel Optimization.}
Most performance optimization benchmarks focus on optimizing the implementation of domain-specific languages for targeted hardware devices.
KernelBench~\citep{kernelbench} asks models to produce functionally correct CUDA kernels that achieve speedup over a PyTorch baseline on NVIDIA GPUs.
TritonBench~\citep{tritonbench} evaluates LLMs on kernel operator generation with hardware-aware performance profiling, assessing both execution accuracy and performance efficiency across deployed GPUs.
While these benchmarks optimize software for fixed hardware, \hsco{} requires the agent to design custom hardware accelerators with system-level considerations, effectively making the hardware itself a part of the design space.

\begin{table}[t]
  \caption{Comparison of our benchmark with related work. }
  \label{tab:comparison}
  \centering
  \scriptsize
  \resizebox{\textwidth}{!}{%
  \begin{tabular}{l c c c c c c c}
    \toprule
    \textbf{Benchmark} & \textbf{Domain} & \shortstack{\textbf{Design}\\\textbf{Entry}} & \shortstack{\textbf{Kernel}\\\textbf{Selection}} & \shortstack{\textbf{HW}\\\textbf{Design}} & \shortstack{\textbf{SoC}\\\textbf{Integ.}} & \shortstack{\textbf{Speedup}\\\textbf{Eval.}} & \shortstack{\textbf{Resource}\\\textbf{Aware}} \\
    \midrule
    \multicolumn{8}{l}{\textit{Software Engineering \& Kernel Optimization}} \\
    KernelBench~\citep{kernelbench}  & SW    & CUDA     & \checkmark   & $\times$   & $\times$   & \checkmark   & $\times$ \\
    TritonBench~\citep{tritonbench}  & SW    & Triton     & $\times$   & $\times$   & $\times$   & \checkmark   & \checkmark \\
    \midrule
    \multicolumn{8}{l}{\textit{RTL \& Chip-Level Hardware Design}} \\
    RTLLM~\citep{rtllm}              & HW    & Verilog  & $\times$   & \checkmark & $\times$   & $\times$   & $\times$   \\
    VerilogEval v2~\citep{verieval_v2} & HW  & Verilog  & $\times$   & \checkmark & $\times$   & $\times$   & $\times$   \\
    ResBench~\citep{resbench}        & HW    & Verilog  & $\times$   & \checkmark & $\times$   & $\times$   & \checkmark   \\
    CVDP~\citep{cvdp}                & HW    & Verilog  & $\times$   & \checkmark & $\times$   & $\times$   & $\times$   \\
    ChipBench~\citep{chipbench}      & HW    & Verilog  & $\times$   & \checkmark & $\times$   & $\times$   & $\times$   \\
    SLDB~\citep{sldb}                & HW &  Verilog  & $\times$   & \checkmark   & \checkmark & $\times$ & $\times$   \\
    
    \midrule
    \multicolumn{8}{l}{\textit{HLS-Based Accelerator Design}} \\
    HLS-Eval~\citep{hls_eval}        & HW    & C/C++    & $\times$   & \checkmark & $\times$   & \checkmark   & $\times$   \\
    Gai et al.~\citep{hls_bench1}    & HW    & C/C++    & $\times$   & \checkmark & $\times$   & $\times$   & $\times$   \\
    HLSYN~\citep{hlsyn}     & HW    & C/C++    & $\times$   & \checkmark   & $\times$   & $\times$   & \checkmark   \\
    
    \midrule
    \textbf{\hsco}                    & \textbf{HW/SW} & \textbf{C + SystemC} & \checkmark & \checkmark & \checkmark & \checkmark & \checkmark \\
    \bottomrule
  \end{tabular}
  }
\end{table}

\textbf{LLM Benchmarks for RTL and Chip-Level Hardware Design.}
Existing benchmarks evaluating LLMs on hardware design tasks typically assume a fixed algorithmic target.
RTLLM~\citep{rtllm} and the updated VerilogEval suite~\citep{verieval_v2} require models to generate Verilog modules from natural language specifications and grade them on functional correctness.
ResBench~\citep{resbench} adds a resource-aware angle by evaluating LLM-generated FPGA designs based on their hardware resource usage.
CVDP~\citep{cvdp} broadens the scope to a comprehensive set of RTL design and verification problems, while ChipBench~\citep{chipbench} extends the evaluation toward larger chip-level designs.
On the other hand, SLDB~\citep{sldb} targets chip-level heterogeneous SoC design, but focuses specifically on measuring the LLM's capability to integrate a pre-defined accelerator into an SoC platform.
These benchmarks have driven rapid progress in hardware-focused design and integration across different scales; yet they operate under the assumption that ``what the hardware should compute'' has already been decided.
Our work augments these efforts by incorporating software execution requirements directly into the hardware design loop, challenging the agent to navigate system-level integration and the hardware-software interface.

\textbf{LLM Benchmarks for HLS-Based Accelerator Design.}
HLS-Eval~\citep{hls_eval} and Gai et al.~\citep{hls_bench1} evaluate LLMs on standalone HLS kernels, measuring whether generated C/C++ or SystemC compiles, synthesizes, and matches a reference.
HLSYN~\citep{hlsyn} provides an FPGA HLS dataset for training quality-of-results predictors.
We share with this line of work the choice of HLS as the hardware design abstraction, but two differences are crucial:
(i)~these benchmarks evaluate accelerators in isolation, whereas \hsco{} requires integration into a full SoC executing a software application;
and (ii)~our agent must autonomously identify which kernel to accelerate, rather than receiving a predefined kernel description.

Table~\ref{tab:comparison} summarizes the comparison.

\section{\hsco{}}
\label{section:bench}
This section presents the design of \hsco{}. We first describe the test case collection process and the codebase structure that underlies the benchmark. We then define the task format, specifying the inputs provided to LLM agents and the expected outputs, along with the evaluation metrics used to assess correctness, performance, and resource utilization. Finally, we discuss the broader impact and extensibility of the benchmark.

\subsection{Testcase Collection and Codebase Structure}
\label{subsec:testcase_collect}

\begin{table}[t]
  \caption{Overview of \hsco{} Application Suite.}
  \label{tab:app}
  \centering
  \scriptsize
  \footnotesize %
  \renewcommand{\arraystretch}{1.1} %
  \setlength{\tabcolsep}{6pt} %
  
  \begin{tabular}{l p{7.2cm} l} %
    \toprule
    \textbf{Name} & \textbf{Description} & \textbf{Sources} \\
    \midrule
    
    \rowcolor[gray]{0.92} \multicolumn{3}{l}{\textbf{Anomaly Detection}} \\
    Autoencoder & FC-based model designed for industrial manufacturing. & \citep{autoencoder_test_case} \\
    
    \midrule
    \rowcolor[gray]{0.92} \multicolumn{3}{l}{\textbf{Natural Language Processing}} \\
    BERT-Tiny  & Distilled Transformer for general sentence classification. & \citep{bert_tiny_1, bert_tiny_2} \\
    MiniLM     & Transformer for paraphrase detection (L3-v2 variant). & \citep{minilm} \\
    TinyCLIP-T & Text encoding component of the TinyCLIP architecture. & \citep{tinyclip} \\
    
    \midrule
    \rowcolor[gray]{0.92} \multicolumn{3}{l}{\textbf{Computer Vision}} \\
    EfficientViT & Low-latency Transformer-based image classifier. & \citep{efficientvit} \\
    MiniDeiT     & Data-efficient Image Transformer (small variant). & \citep{MiniViT} \\
    TinyCLIP-V   & Vision component of TinyCLIP (reduced to 1-layer). & \citep{tinyclip} \\
    MobileNetV4  & CNN optimized for mobile deployment efficiency. & \citep{mobilenetv4_1, mobilenetv4_2} \\
    
    \midrule
    \rowcolor[gray]{0.92} \multicolumn{3}{l}{\textbf{Signal Processing}} \\
    NightVision  & Low-light contrast improvement via Histogram Equalization. & \citep{PERFECT} \\
    WAMI         & ISP pipeline: debayering, Lucas-Kanade, change detection. & \citep{PERFECT, porter_spmag2010} \\
    \bottomrule
  \end{tabular}
\end{table}

\textbf{Testcase Collection.}
We curated a diverse set of 10 workloads from open-source GitHub repositories spanning machine learning and digital signal processing (Table~\ref{tab:app}). These applications were ported to a RISC-V bare-metal environment using ESP-compatible function calls. Non-C programs were translated into C (with LLM assistance during data curation) and manually verified. To accommodate the slow physical execution speed of FPGAs, the workloads were carefully scaled to complete within one hour of bare-metal evaluation on a Virtex-7 VC707 FPGA (details in Appendix~\ref{app:runtime}).

\textbf{Codebase Structure.}
The benchmark codebase is logically partitioned into three distinct parts (comprehensive directory tree in Appendix~\ref{app:codebase}). 
For the \emph{software part}, \texttt{systest.c} serves as the entry point, relying on unified headers (\texttt{accel\_common.h}, \texttt{accel\_drivers.h}) for standardized accelerator registration and operations. 
For the \emph{accelerator part}, components are encapsulated within a dedicated \texttt{<accelerator>\_workspace}. This allows agents to generate new hardware by cloning a template and modifying the localized files such as SystemC implementations, register interfaces, and software drivers, guided by the provided \texttt{README.md} instructions. 
Finally, for the \emph{system integration part}, the target SoC features a tile-based architecture pre-configured with one CPU tile, one memory tile for DRAM interaction, and one I/O tile for peripheral access. Agents integrate their accelerators by updating the SoC configuration file, while deployment scripts handle data transfer to the FPGA.

\subsection{Task Format and Evaluation Metrics}
\label{subsec:task_format}

\begin{figure}[t]
\begin{center}
\includegraphics[width=0.9\columnwidth]{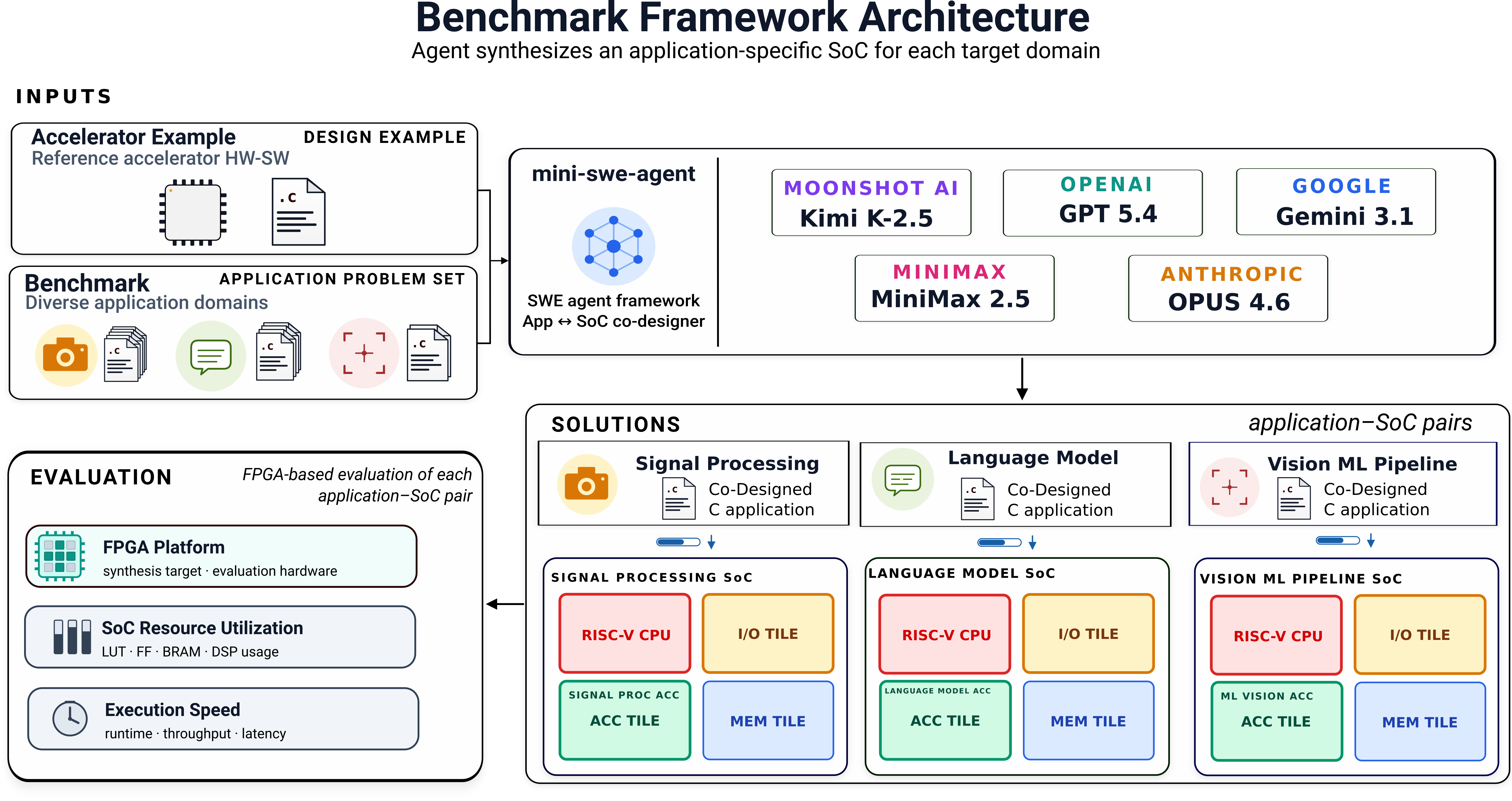}
\caption{\hsco{} flow.}
\label{fig:bench_flow}
\end{center}
\vspace{-15pt}
\end{figure}

\figurename~\ref{fig:bench_flow} shows the HSCO-Bench flow. For each problem, the LLM agent is tasked with the following:

\textbf{Input:} The problem set consists of two primary components.
First, the agent is provided with an accelerator template to serve as a reference for designing HLS-synthesizable accelerators.
Second, each problem folder contains the original C application code and a comprehensive \texttt{README.md} file. This file explains the codebase structure, provides clear instructions for implementation and SoC integration, and specifies the software driver interface required to offload computation to the hardware accelerator.
The full content of the \texttt{README.md} is provided in Appendix~\ref{app:readme}.

\textbf{Output:} For each problem, the LLM agent is expected to autonomously identify potential performance bottlenecks and generate the solution by modifying files in the problem folder:
\begin{enumerate}
    \item \textbf{Hardware Generation:} Design one or more custom accelerators following the provided template, generating the synthesizable SystemC implementations along with their corresponding software drivers.
    \item \textbf{Software Modification:} Replace relevant parts of the original application code with driver calls to offload operations to the accelerator. This process may also involve optimizing data movement and execution strategies to improve system throughput.
    \item \textbf{System Integration:} Update the SoC configuration files to integrate the newly generated accelerators into the target heterogeneous platform.
\end{enumerate}

The generated outputs, including the modified application code, hardware kernels, and SoC configuration, are subsequently deployed to the FPGA board for validation.

\textbf{Evaluation Metrics:}
We evaluate the outputs generated by the LLM agents based on three metrics:

\begin{enumerate}
    \item \textbf{Correctness:} Application execution outputs are compared against golden references generated from the original software. We define a tolerance range to accommodate potential minor discrepancies introduced by hardware-specific optimizations or precision differences (e.g., floating-point versus fixed-point arithmetic) during accelerated execution.
    \item \textbf{Performance.} Speedup is the ratio of CPU cycles for the software-only baseline ($C_{\text{sw}}$) to those for the accelerated execution ($C_{\text{acc}}$): $S = C_{\text{sw}} / C_{\text{acc}}$. Both are measured directly on the RISC-V CPU at runtime; cycle-counting details are provided in Appendix~\ref{app:cycle}.
    \item \textbf{Resource Utilization:} We monitor three primary hardware resource categories on the FPGA: Logic Cells, BRAMs, and DSP slices. In Section~\ref{subsec:exp_overall_perf}, we report the average percentage usage of these resources to characterize the hardware overhead. 
\end{enumerate}

Performance and resource utilization are orthogonal dimensions of the evaluation. In our experiments, we instruct the LLM agents to maximize speedup under the constraint that the design fits within the available resources of the target Virtex-7 VC707 FPGA board. This reflects a common design goal where performance is prioritized as long as the resource budget is not exceeded.

The evaluation scripts and methodologies are detailed in Appendix~\ref{app:scripts}.

\subsection{Impact and Extensibility}
\label{subsec:impact}

The interplay between performance speedup and hardware resource utilization defines a complex design space. While our current evaluation focuses on maximizing speedup using available FPGA resources, the underlying infrastructure of \hsco{} is inherently multidimensional. The benchmark can be easily extended by imposing strict resource ceilings or performance requirements, challenging agentic frameworks to intelligently navigate performance-resource trade-offs along the Pareto frontier.

Furthermore, the benchmark is highly extensible in both its software workloads and physical deployment. On the software side, the modular workspace and standardized integration interfaces encourage community-driven contributions, allowing researchers to easily append new applications and testcases to the suite. On the hardware side, leveraging extensible open-source frameworks---with ESP serving as our foundational vehicle---extends the benchmark's relevance all the way to physical silicon. Because the underlying SoC architecture and integration scripts are silicon-proven~\citep{epochs1_isscc}, \hsco{} offers a realistic, low-risk pathway from LLM-generated logic to real-world ASIC implementation (e.g., GDSII generation).

We envision \hsco{} serving as a catalyst for the research community. By providing a standardized framework on an open-source SoC platform for evaluating LLM-driven hardware-software co-design, it facilitates systematic progress in complex system-level automation.

\section{Experimental Results}
\label{section:results}

\textbf{Experimental Setup.}
We evaluate \hsco{} using the \texttt{mini-swe-agent}~\citep{swe_agent} (version 2.2.8) across five LLMs: MiniMax-M2.5~\citep{minimax_m25} and Kimi-K2.5~\citep{kimi_k_25} via DeepInfra~\citep{deepinfra}, alongside Claude Opus 4.6~\citep{opus_4_6}, GPT-5.4~\citep{gpt_5_4}, and Gemini 3.1 Pro~\citep{gemini_3_1_pro} via their official APIs. Accelerator implementations are synthesized using Catapult HLS~\citep{catapult_hls} on an AMD Ryzen 9 9950X machine (192GB RAM) and validated on an AMD Xilinx Virtex-7 XC7VX485T FPGA.

\subsection{Overall Performance and Capability Gap}
\label{subsec:exp_overall_perf}

\begin{table}[t]
  \caption{Performance and resource utilization of generated SoCs across different models.}
  \label{tab:agent_comparison}
  \centering
  \scriptsize
  \renewcommand{\arraystretch}{1.2}
  
  \begin{tabularx}{\columnwidth}{l l *{5}{>{\centering\arraybackslash}X} r}
    \toprule
    \textbf{Application} & \textbf{Metric} & \textbf{MiniMax M2.5} & \textbf{Kimi K2.5} & \textbf{Opus 4.6} & \textbf{GPT-5.4} & \textbf{Gemini 3.1 Pro} & \textbf{CPU Baseline} \\
    \midrule
    \multirow{2}{*}{Autoencoder} & Speedup  & --- & --- & 1.01$\times$ & 1.00$\times$ & --- & $1.4 \times 10^8$ \\
                                 & Resource & --- & --- & +4.04\% & +0.0\% & --- & (Base 13.02\%) \\
    \midrule
    \multirow{2}{*}{BERT-Tiny}   & Speedup  & --- & --- & 5.18$\times$ & 1.00$\times$ & --- & $6.1 \times 10^9$ \\
                                 & Resource & --- & --- & +2.38\% & +2.13\% & --- & (Base 13.02\%) \\
    \midrule
    \multirow{2}{*}{EfficientViT}& Speedup  & --- & --- & 3.03$\times$ & --- & --- & $1.0 \times 10^{10}$ \\
                                 & Resource & --- & --- & +2.37\% & --- & --- & (Base 13.02\%) \\
    \midrule
    \multirow{2}{*}{MiniDeiT}    & Speedup  & --- & --- & 3.59$\times$ & 3.72$\times$ & --- & $1.3 \times 10^{11}$ \\
                                 & Resource & --- & --- & +4.98\% & +17.45\% & --- & (Base 13.02\%) \\
    \midrule
    \multirow{2}{*}{MiniLM}      & Speedup  & 1.00$\times$ & --- & 9.69$\times$ & --- & --- & $7.3 \times 10^{10}$ \\
                                 & Resource & +0.0\% & --- & +2.36\% & --- & --- & (Base 13.02\%) \\
    \midrule
    \multirow{2}{*}{MobileNetV4} & Speedup  & --- & --- & --- & --- & --- & $2.6 \times 10^{10}$ \\
                                 & Resource & --- & --- & --- & --- & --- & (Base 13.02\%) \\
    \midrule
    \multirow{2}{*}{TinyCLIP-T}  & Speedup  & --- & --- & 7.44$\times$ & 16.22$\times$ & 1.00$\times$ & $1.9 \times 10^{10}$ \\
                                 & Resource & --- & --- & +4.98\% & +15.37\% & +0.0\% & (Base 13.02\%) \\
    \midrule
    \multirow{2}{*}{TinyCLIP-V}  & Speedup  & --- & --- & --- & 11.02$\times$ & --- & $1.8 \times 10^{10}$ \\
                                 & Resource & --- & --- & --- & +23.67\% & --- & (Base 13.02\%) \\
    \midrule
    \multirow{2}{*}{NightVision} & Speedup  & --- & --- & 1.08$\times$ & --- & --- & $2.2 \times 10^8$ \\
                                 & Resource & --- & --- & +1.12\% & --- & --- & (Base 13.02\%) \\
    \midrule
    \multirow{2}{*}{WAMI}        & Speedup  & --- & --- & --- & --- & --- & $4.1 \times 10^8$ \\
                                 & Resource & --- & --- & --- & --- & --- & (Base 13.02\%) \\
    \midrule[1pt]
    \multirow{3}{*}{\textbf{Summary}} & \textbf{Avg. Spd} & 1.00$\times$ & 1.00$\times$ & \textbf{2.32$\times$} & 1.92$\times$ & 1.00$\times$ & --- \\
                                      & \textbf{Avg. Res} & +0.0\% & +0.0\% & +2.22\% & +5.86\% & +0.0\% & --- \\
                                      & \textbf{Success}  & 10\% & 0\% & \textbf{70\%} & 50\% & 10\% & --- \\
    \bottomrule
  \end{tabularx}
\end{table}

\begin{figure}[t]
     \centering
     \begin{subfigure}[b]{0.49\columnwidth}
         \centering
         \includegraphics[width=\linewidth]{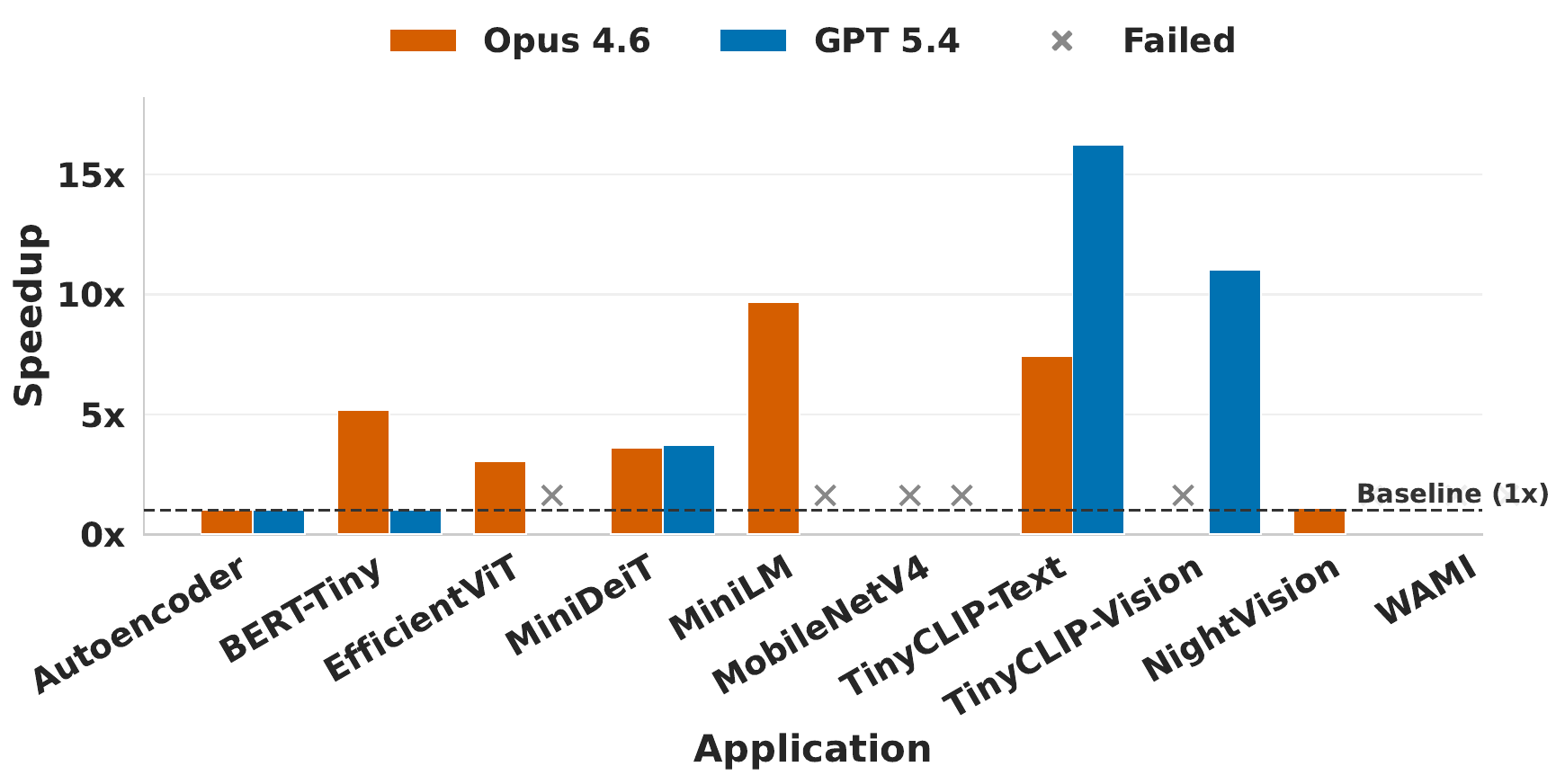}
         \caption{Speedup}
         \label{fig:speedup}
     \end{subfigure}
     \hfill
     \begin{subfigure}[b]{0.49\columnwidth}
         \centering
         \includegraphics[width=\linewidth]{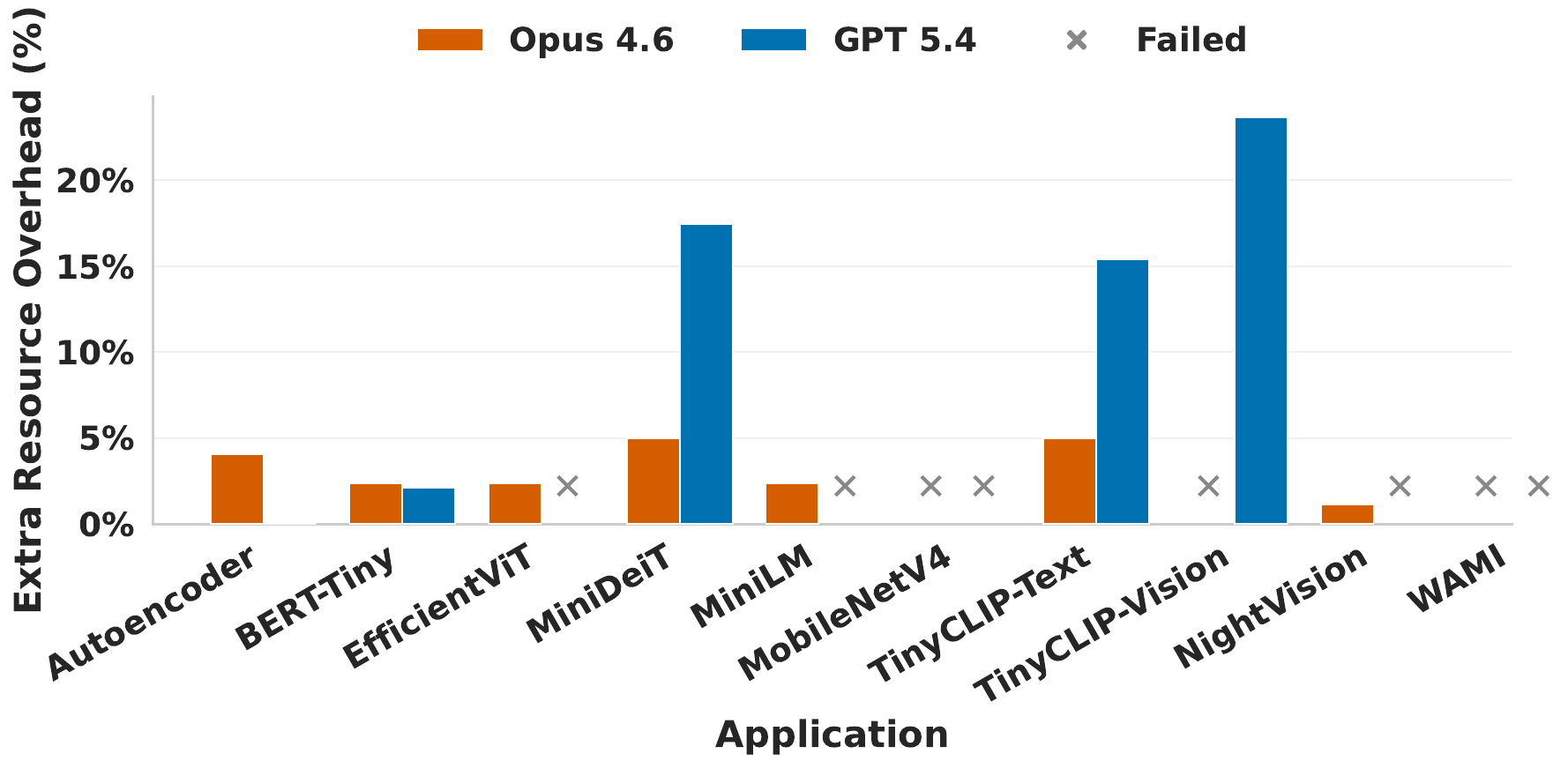}
         \caption{Resource Overhead}
         \label{fig:resource}
     \end{subfigure}
     \caption{Evaluation results across 10 applications.}
     \label{fig:evaluation_results}
     \vspace{-15pt}
\end{figure}

The benchmark requires simultaneous reasoning over hardware synthesis and software execution logic. Table~\ref{tab:agent_comparison} details the performance and resource usage across all testcase-model pairs. To standardize the evaluation, speedups are reported relative to bare-metal CPU execution, and resource utilization represents the overhead beyond the base SoC's inherent 13.02\% usage. Average speedup is calculated as a geometric mean, while average resource overhead uses an arithmetic mean across key FPGA resources (failed attempts are conservatively assigned a $1\times$ speedup and $+0\%$ overhead). The total cost of the experiment is provided in Appendix~\ref{app:total-cost}.

The results reveal a capability gap: \textbf{Claude Opus 4.6} and \textbf{GPT-5.4} are the only models capable of meaningfully engaging with the end-to-end co-design tasks. Conversely, other models almost entirely failed to navigate the complex SoC integration process, with nearly a 0\% success rate.

\figurename~\ref{fig:evaluation_results} visualizes the performance dynamics between the two capable frontier models. Opus 4.6 delivered the most robust overall performance, achieving a 70\% success rate and a $2.32\times$ average speedup with a remarkably low average resource footprint (+2.22\%). Although GPT-5.4 recorded a lower success rate (50\%) and average speedup ($1.92\times$), it exhibited the highest \textit{peak} performance. For instance, in the TinyCLIP-T and TinyCLIP-V applications, GPT-5.4 achieved massive speedups of $16.22\times$ and $11.02\times$, at the cost of higher resource utilization (+15.37\% and +23.67\%, respectively).

Based on the empirical results in Table~\ref{tab:agent_comparison}, we categorize the testcases into three empirical success tiers to illustrate the difficulty gradient of our benchmark:
\begin{enumerate}[leftmargin=*]
    \item \textbf{High Success:} Cases where both frontier models succeed, namely Autoencoder, BERT-Tiny, MiniDeiT, and TinyCLIP-T.
    \item \textbf{Mixed Success:} Cases where only one model succeeds, including EfficientViT, MiniLM, TinyCLIP-V, and NightVision.
    \item \textbf{Consistent Failure:} Cases where both models fail, specifically MobileNetV4 and WAMI.
\end{enumerate}

\subsection{Error Profiling and Integration Bottlenecks}
\label{subsec:exp_bottlenecks}

\begin{wrapfigure}{r}{0.5\textwidth}
    \centering
    \vspace{-10pt}
    \includegraphics[width=\linewidth]{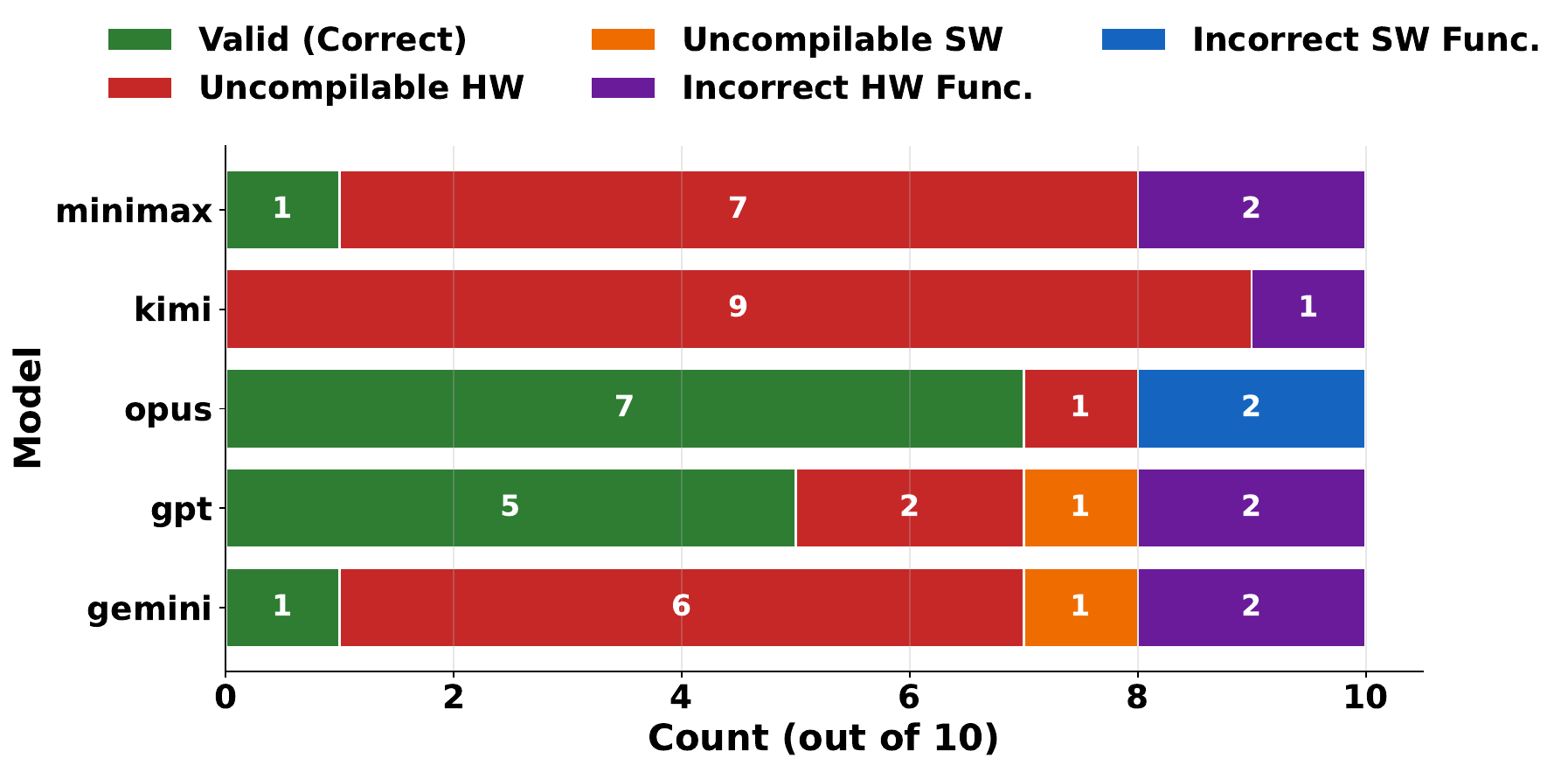}
    \caption{Distribution of failure modes across evaluated LLM agents.}
    \label{fig:error}
\end{wrapfigure}

To understand the widespread failures among most models, we categorize the integration bottlenecks into four primary modes (\figurename~\ref{fig:error}): (1) \textbf{Uncompilable Hardware} (invalid SystemC syntax or HLS synthesis failures); (2) \textbf{Uncompilable Software} (C syntax errors or bare-metal violations); (3) \textbf{Incorrect Hardware Functionality} (failing behavioral simulation); and (4) \textbf{Incorrect Software Functionality} (runtime hangs or numerical mismatches). Analysis of these modes reveals key system-level bottlenecks:

\textbf{Syntax and Interface Inconsistencies:} The most frequent barrier for models like MiniMax-M2.5 and Gemini 3.1 Pro is maintaining pipeline consistency. MiniMax consistently generated incomplete HLS kernels missing standard register interfaces. Similarly, Gemini often neglected to update hardware device IDs and produced misaligned testbenches that failed to match its own custom accelerator interfaces, preventing designs from reaching synthesis.

\textbf{Over-ambitious Design and the Multi-Accelerator Bottleneck:} Interestingly, some models often failed due to overly ambitious architectural choices. Kimi-K2.5 frequently ignored provided templates to build entire subsystems from scratch. For BERT-Tiny, it even attempted to concurrently instantiate independent GELU, LayerNorm, and MatMul units. However, orchestrating multi-accelerator designs and refining the hardware-software interfaces for complex SoC configurations overwhelmed the model, universally resulting in massive code syntax errors and integration failures.

\textbf{Software and Bare-Metal Environment Gaps:} Even capable models faced friction at the hardware-software boundary. GPT-5.4 struggled with Direct Memory Access (DMA) logic and data alignment, failing the test during the behavioral simulation. It also occasionally hallucinated standard library includes that are unsupported in the specialized bare-metal environment, leading to uncompilable software. Meanwhile, Claude Opus 4.6's rare pipeline failures were predominantly late-stage issues, such as software hangs or numerical mismatches during execution.

\subsection{Micro-Architectural Insights and Case Studies}
\label{subsec:exp_case_studies}

\textbf{Micro-Architectural Trade-offs:} Opus 4.6 and GPT-5.4 exhibit diverging design philosophies when tackling the same workloads. Opus 4.6 frequently designs \textit{small-footprint accelerators} and relies heavily on \textit{software-managed tiling} to process machine learning kernels (e.g., in \textbf{BERT-Tiny}, Appendix~\ref{app:bert_opus}). This strategy keeps hardware resource utilization strictly bounded but introduces software overhead due to frequent CPU-accelerator data transfers. Conversely, GPT-5.4 favors \textit{monolithic, large-scale accelerators}. While utilizing more hardware resources, this approach yields superior peak performance for applications like \textbf{TinyCLIP} (\figurename~\ref{fig:evaluation_results}) by effectively bypassing complex software data-orchestration bottlenecks.

\textbf{Operator Fusion and Domain-Specific Extraction:} Beyond simple offloading, frontier models demonstrated the capability to perform more sophisticated operator fusion. For the \textbf{Autoencoder}, Opus 4.6 correctly identified the \texttt{dense\_bn\_relu} sequence as a cohesive logical block. Rather than offloading operations individually, it fused the matrix multiplication (\texttt{dense}) and ReLU into a single custom HLS accelerator, while strategically delegating the batch normalization (\texttt{bn}) layer to the host CPU (Appendix~\ref{app:autoencoder_opus}). Furthermore, while agents predominantly target standard GEMMs, they can occasionally extract non-standard logic: in \textbf{NightVision}, Opus 4.6 successfully synthesized a sorting-based \texttt{slowMedian3x3} kernel (Appendix~\ref{app:nightvision_opus}). This observation highlights the versatility of LLMs, demonstrating their potential to move beyond predefined templates and generate customized hardware accelerators purely driven by specific software needs.

\paragraph{Conservative Design and Hardware Underutilization.}
Although strong frontier models successfully achieve functional correctness, we observed a prevalent conservative behavior: they almost exclusively instantiate SoCs with a single common accelerator, such as a matrix multiplier. Conversely, while models like Kimi-K2.5 attempted to generate more ambitious multi-accelerator designs, these attempts ultimately failed during hardware validation. This reveals a critical optimization gap. In real-world engineering, multi-accelerator integration and sophisticated data transfer strategies are essential for maximizing system throughput and hardware utilization, which is a complex task that currently still requires months of manual human expert effort.

\subsection{Cost Efficiency}
\label{subsec:cost_eff}

\begin{wrapfigure}{r}{0.5\textwidth} %
    \centering
    \vspace{-10pt}
    \includegraphics[width=\linewidth]{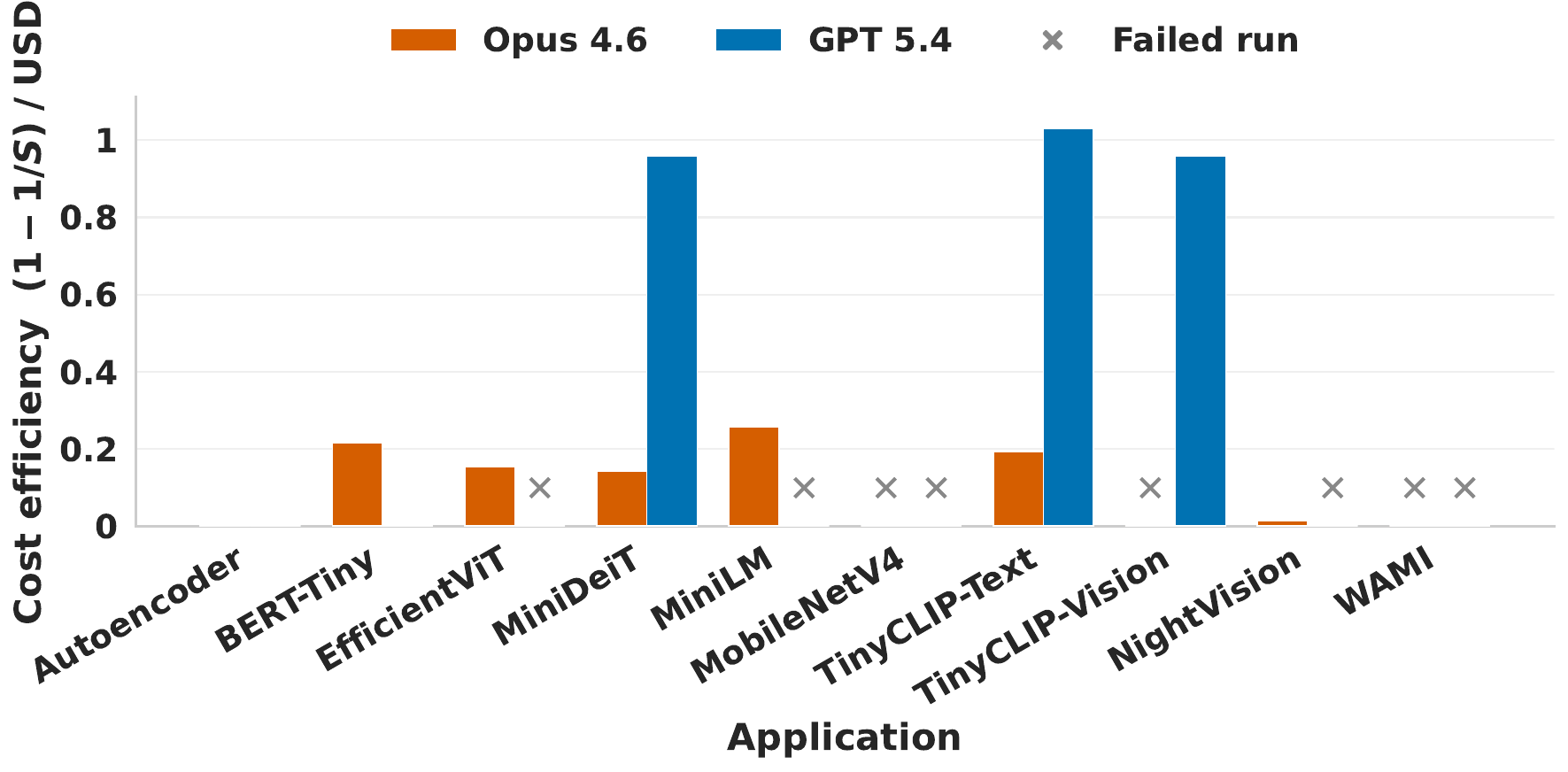}
    \caption{Cost efficiency ($\eta$) comparison across applications. While Opus 4.6 provides a steady baseline with fewer failures, GPT-5.4 exhibits superior value-for-money on several successful runs.}
    \label{fig:speedup_per_dollar}
\end{wrapfigure}

To evaluate the economic viability of utilizing LLMs for SoC design, we assess the cost efficiency of the two successful models (Opus 4.6 and GPT-5.4). We define a specialized metric, \textit{Cost Efficiency} ($\eta$), calculated as:\begin{equation}\eta = \frac{1 - \frac{1}{S}}{C},\end{equation}where $S$ represents the achieved speedup and $C$ denotes the API cost in USD for the attempt. The numerator, $1 - 1/S$, quantifies the percentage of execution time saved compared to the baseline CPU execution.
As illustrated in \figurename~\ref{fig:speedup_per_dollar}, GPT-5.4 achieved a $7.14\times$ higher cost efficiency than Opus 4.6 on average. While Opus 4.6 delivered a higher success rate and superior overall speedup, the elevated inference costs associated with its multi-step reasoning yielded lower marginal returns. In contrast, GPT-5.4 provided a better balance between design complexity and performance gains, delivering a superior cost-to-performance ratio across several benchmark tasks.

\section{Conclusion}

In this work, we introduced \hsco{}, the first end-to-end hardware-software co-design benchmark for agent-driven heterogeneous SoC generation.
Built upon the ESP open-source SoC platform, \hsco{} evaluates whether frontier LLMs can analyze software code, autonomously generate accelerators, integrate them into an SoC, and verify functionality.
Our results show that fully autonomous SoC generation remains challenging, yet state-of-the-art agents can successfully construct valid SoC prototypes with measurable speedups.
However, hardware resource utilization remains low, indicating there is substantial room for improved design strategies and optimization.
Furthermore, \hsco{} provides a highly scalable foundation in three directions:
community-driven testcase expansion through its modular format,
extension toward broader hardware-software design challenges on a silicon-proven platform,
and future studies exploring performance-resource trade-offs under varying resource constraints.
We believe \hsco{} paves the way toward fully autonomous hardware-software co-design, potentially enabling AI-generated customized SoCs for self-acceleration.

\bibliographystyle{plainnat} 
\bibliography{main-ref}

\appendix

\newpage 
\section*{Appendices}

\section{Execution Time of Applications on the FPGA}
\label{app:runtime}

\begin{table}[h]
  \caption{CPU runtime of each workload on the FPGA.}
  \label{tab:runtime}
  \centering
  \begin{tabular}{l r} %
    \toprule
    Name             & runtime \\
    \midrule
    Autoencoder      & 33s     \\
    BERT-Tiny       & 3m 06s  \\
    EfficientViT     & 4m 11s  \\
    MiniDeiT         & 44m 50s \\
    MiniLM           & 25m 36s \\
    MobileNetV4      & 9m 45s  \\
    TinyCLIP-T   & 7m 21s  \\
    TinyCLIP-V & 7m 42s  \\
    NightVision      & 32s     \\
    WAMI             & 36s     \\
    \bottomrule
  \end{tabular}
\end{table}

\tablename~\ref{tab:runtime} reports the execution time of each application when deployed on a CPU-only SoC and emulated on the Virtex-7 FPGA VC707 Evaluation Kit.
The total execution time includes the time required to:
\begin{enumerate}
    \item load the bitstream onto the FPGA
    \item load the data onto the FPGA
    \item compile the program
    \item execute the program
\end{enumerate}

\section{Codebase Structure of Each Testcase}
\label{app:codebase}

Each test case is located in a separate directory, and the files are listed in \figurename~\ref{fig:code_structure}.

\begin{figure}[h]
\centering
\begin{minipage}{0.95\textwidth} 
\dirtree{%
.1 test\_case/.
.2 <accelerator>\_workspace/ \DTcomment{accelerator related folder for each accelerator, including accelerator systemC code, simulation scripts, and software driver}. 
.2 data\_files/ \DTcomment{data required by applications}.
.2 accel\_common.h \DTcomment{common functions for accelerator utility}.
.2 accel\_drivers.h \DTcomment{entry point for accelerators to register}.
.2 esp\_xilinx-vc707-xc7vx485t\_defconfig \DTcomment{SoC configuration file}.
.2 load\_data\_ariane.sh \DTcomment{data loading script to load data to FPGA}.
.2 Makefile \DTcomment{application compilation and execution commands}.
.2 systest.c \DTcomment{application entry point: main()}.
.2 xxx.c, xxx.h \DTcomment{source code for applications}.
.2 testbench.vhd \DTcomment{testbench of SoC}.
.2 top.vhd \DTcomment{top-level SoC verilog}.
}
\end{minipage}
\caption{The testcase structure of our proposed benchmark.}
\label{fig:code_structure}
\end{figure}

\section{Prompt and README for mini-swe-agent}
\label{app:readme}
The prompt and README used in each testcase.

\begin{promptbox}{Prompt entered into mini-swe-agent}
    "Please take a look at the README.md file and finish the job."
\end{promptbox}

\begin{markdownprompt}{Instruction for SoC Generation Flow (README.md)}
# FPGA Accelerator Design - ESP / VC707 / Ariane Baremetal

## Project Overview

This workspace contains a compute-intensive workload targeting the **Xilinx VC707 FPGA board** (xc7vx485t) with an **Ariane RISC-V processor**, running in **baremetal** mode (no OS, no libc malloc - uses `aligned_malloc`). Your task is to **analyze the workload in this workspace**, **design hardware accelerators** and their **software drivers** to speed up its execution, then **integrate them into the existing software**. Workloads vary across testcases - some are ML inference (autoencoder, transformer, CNN), others are classical signal / image processing (image fusion, motion imagery, etc.). The exact workload and which operations dominate runtime are **for you to discover** by reading the source files.

## File Structure

### Required Accelerator Directory Layout (strict)

Every accelerator you build MUST live at exactly this path inside the workspace:

```
<name>_workspace/
├── <name>_sysc_catapult/
│   ├── <name>_sysc_catapult.xml
│   ├── src/<name>.cpp
│   ├── inc/
│   │   ├── <name>.hpp
│   │   ├── <name>_specs.hpp
│   │   ├── <name>_conf_info.hpp
│   │   └── <name>_data_types.hpp
│   ├── tb/
│   │   ├── testbench.cpp
│   │   └── testbench.hpp
│   ├── hls/                       ← validator runs `make` here; MUST contain:
│   │   ├── Makefile               ← REQUIRED - defines bev-sim, hls, install-force targets
│   │   ├── build_prj.tcl
│   │   ├── build_prj_top.tcl
│   │   ├── common.tcl
│   │   └── rtl_sim.tcl
│   └── hls-work-virtex7/          (build output - can start empty, populated by `make hls`)
└── <name>_sw_driver/
    ├── <name>_cfg.h
    └── <name>_driver.h
```

The outer `<name>_workspace/` wrapper is **required**. The `hls/` directory must contain the **full set of build scripts** copied from `dummy_workspace/dummy_sysc_catapult/hls/` - an empty `hls/` (even with `.gitkeep`) will satisfy the existence check but fail at `make bev-sim-clean` with "No rule to make target". `check_workspace.sh` reads accelerator names from `esp_xilinx-vc707-xc7vx485t_defconfig` and runs `make bev-sim && make hls && make install-force` inside `<name>_workspace/<name>_sysc_catapult/hls/`. Mirror the reference `dummy_workspace/` exactly - the simplest way is `cp -r dummy_workspace <name>_workspace` followed by a rename pass (see Step 2).

### Software Files (Baremetal C)

| File | Description |
|------|-------------|
| `systest.c` | Main entry point. Loads data, allocates buffers, runs the workload on test samples, measures execution time. Contains the `USE_ACCELERATOR` toggle. |
| `<workload>.h` (and optionally `<workload>.c`) | Core computation invoked from `systest.c`. For ML workloads this is the forward pass; for signal / image-processing workloads it is the main algorithm. Exact filename depends on the workload - inspect the workspace to find it. |
| `nn_ops.h` *(ML workloads only, if present)* | Shared neural-network operation primitives used by the forward pass. Contents depend on the model; inspect to see which ops (linear, conv, normalization, activation, etc.) exist. Non-ML workloads (image processing, etc.) will not have this file; their ops live directly inside `<workload>.c`. |
| `tensor.h` *(ML workloads only, if present)* | Tensor struct definition, model loading from memory-mapped binary, memory allocation utilities. Non-ML workloads may omit this file and load raw input / gold-output buffers directly. |
| `accel_drivers.h` | Accelerator driver registry - single include point for all accelerator drivers. When adding a new accelerator: (1) `#include` its driver header, (2) add init call in `accel_init_all()`, (3) add cleanup call in `accel_cleanup_all()`. |
| `accel_common.h` | Shared accelerator infrastructure: DMA scatter/gather macros, shared memory allocation (`accel_shared_mem`, 512KB), fixed-point conversion (`float2fixed`/`fixed2float`). Included once by `accel_drivers.h`. |

### Accelerator Workspace (`dummy_workspace/`)

This directory contains a **structural reference accelerator** that demonstrates the shape of an ESP accelerator workspace: HLS kernel source, testbench, XML descriptor, and baremetal software driver. The reference kernel is intentionally an **identity pass-through** (it copies input A to output O; input B is loaded but ignored) and performs no useful computation on its own. **You must create your own accelerator workspace(s)** by copying and adapting this reference. The name `dummy` was chosen deliberately so that a plain `sed 's/dummy/myaccel/g'` rename cannot collide with any matchlib or Catapult header path.

| Path | Description |
|------|-------------|
| `dummy_sysc_catapult/` | Reference HLS accelerator (Catapult HLS, SystemC). **Do NOT use this directly** - copy/adapt it. |
| `dummy_sysc_catapult/dummy_sysc_catapult.xml` | Accelerator descriptor: device name, device ID (`200` = `0x200`), register parameters. **Each accelerator must have a unique device_id.** |
| `dummy_sysc_catapult/src/dummy.cpp` | HLS kernel: `config()` → `load()` → `compute()` → `store()` pipeline. `load()` reads A,B via DMA into PLMs; `compute()` is the identity placeholder; `store()` writes O via DMA. |
| `dummy_sysc_catapult/inc/` | Headers: `dummy_specs.hpp` (PLM sizes, DMA config), `dummy_conf_info.hpp` (register params struct), `dummy_data_types.hpp`, `dummy.hpp` (top-level module). |
| `dummy_sysc_catapult/tb/` | SystemC testbench: `testbench.cpp` generates random data, invokes accelerator, verifies against a golden model. |
| `dummy_sysc_catapult/hls-work-virtex7/` | HLS build directory. |
| `dummy_sw_driver/dummy_driver.h` | Baremetal software driver: `dummy_accel_init()` (probe device), `dummy_accel_exec()` (load data → configure regs → start → poll → read results), `dummy_accel_cleanup()`. |
| `dummy_sw_driver/dummy_cfg.h` | Device ID, register offsets (derived from XML), fixed-point bit-width config. |

### Hardware Design Files

| File | Description |
|------|-------------|
| `top.vhd` | VHDL top-level design for the ESP SoC on VC707. |
| `testbench.vhd` | VHDL simulation testbench. |
| `Makefile` | ESP build system for the SoC and software. |
| `esp_xilinx-vc707-xc7vx485t_defconfig` | ESP SoC platform configuration. Defines the NoC grid dimensions, tile layout, and accelerator placement. See **SoC Tile Configuration** below. |

### Data Files

| File | Description |
|------|-------------|
| `<workload>.bin` *(ML workloads)* | Pre-built model binary. Format: magic `0x4145544E` + version + tensors (name, shape, float data). Non-ML workloads may have **no** model binary at all - they load only test input + gold-reference output. |
| `model_files/` *(or `data_files/`)* | Source binaries: model weights (if applicable), test input data, and test labels / gold-reference outputs. Inspect the directory to see the exact files and their sizes. |
| `load_model_ariane.sh` *(or `load_data_ariane.sh`)* | Script to load binary data into FPGA memory via `esplink`. |
| `prepare_data.py` | Python utility to prepare / align the workload's binaries and test data. |

## Accelerator Design Flow
Steps 1-5 in order. Each step depends on the previous one; do not skip ahead or reorder.
### Step 1: Analyze the workload

Start from `systest.c` and trace the execution call graph into the workload-specific header/source (and any shared ops file, if present). Identify the core computations, their sizes, and which operations dominate runtime - those are your acceleration candidates.

### Step 2: Create your accelerator workspace

For each accelerator you want to build, **duplicate the entire `dummy_workspace/` directory** and replace the `dummy`/`DUMMY` name with your accelerator's name (lowercase and UPPERCASE forms). Replace `myaccel` / `MYACCEL` below with your chosen name:

> **Common mistake - DO NOT hand-create the directory tree.** You MUST start with `cp -r dummy_workspace <name>_workspace` (step 1 below). This copies the `hls/` build Makefile and TCL scripts along with all other source files. **An empty `hls/` directory - or one with only `.gitkeep` - will fail** `make bev-sim-clean` with `No rule to make target 'bev-sim-clean'`. Do NOT manually create `<name>_workspace/` or any of its subdirectories, and do NOT place `<name>_sysc_catapult/` directly at the workspace top level without the outer `<name>_workspace/` wrapper. After step 1, verify with: `ls <name>_workspace/<name>_sysc_catapult/hls/Makefile` - if that file is missing, redo the copy.

1. **Copy the workspace and replace `dummy`/`DUMMY` strings inside files** (case-preserving):
```bash
cp -r dummy_workspace myaccel_workspace
cd myaccel_workspace
find . -type f -exec sed -i 's/dummy/myaccel/g; s/DUMMY/MYACCEL/g' {} +
```

2. **Rename files and directories** whose names contain `dummy` - only substitute the basename, not the full path (use `-depth` so inner paths are renamed before their parents):
```bash
find . -depth -name '*dummy*' | while read f; do
  d=$(dirname "$f"); b=$(basename "$f")
  mv "$f" "$d/$(echo "$b" | sed 's/dummy/myaccel/g')"
done
```

3. **Assign a unique device ID** in both the XML file and the driver config header. Pick a value from this project's reserved range `0x201`–`0x27F` (the reference `dummy` already occupies `0x200`). See **Device ID Allocation** below for the full convention.
```bash
sed -i 's/device_id="200"/device_id="201"/' \
  myaccel_sysc_catapult/myaccel_sysc_catapult.xml
sed -i 's/0x200/0x201/g' myaccel_sw_driver/myaccel_cfg.h
```

After these three steps, your new workspace is a functional (but still identity-behavior) clone of the `dummy` reference. You can now redesign the kernel.

### Step 3: Design the accelerator kernel

1. Modify `src/<accel>.cpp` - implement your kernel in the `compute()` function. The `load()` and `store()` functions handle DMA data movement.
2. Modify `inc/<accel>_specs.hpp` - adjust PLM sizes (e.g. `A_PLM_IN_WORD`, `B_PLM_IN_WORD`, `O_PLM_OUT_WORD`) and `MEM_SIZE` to match your data dimensions.
3. If you change the accelerator's configuration parameters, update `inc/<accel>_conf_info.hpp` **and** the XML `<param>` list (register offsets are derived from the XML).
4. Update the testbench (`tb/testbench.cpp`, `tb/testbench.hpp`) to match your accelerator's interface.

### Step 4: Write the software driver

Adapt the driver (from `dummy_sw_driver/dummy_driver.h`) so that it:
1. Probes for the accelerator via `probe(&devs, VENDOR_SLD, DEV_ID, DEV_NAME)`
2. Provides an `exec()` function that:
   - Converts float inputs to fixed-point and writes to `accel_shared_mem`
   - Configures accelerator registers via `iowrite32()`
   - Starts accelerator and polls `STATUS_REG` for completion
   - Converts fixed-point outputs back to float
3. The driver must use the shared memory from `accel_common.h` (do not allocate separate DMA buffers)

### Step 5: Integrate into software

1. Add `#include` for your driver in `accel_drivers.h`
2. Add init/cleanup calls in `accel_init_all()` / `accel_cleanup_all()`
3. Modify the workload's core computation (the file called from `systest.c`) to invoke your accelerator's `exec()` function for the appropriate operations (guarded by `#ifdef USE_ACCELERATOR`)
4. Enable `USE_ACCELERATOR` in `systest.c`
5. Register your accelerator(s) in `esp_xilinx-vc707-xc7vx485t_defconfig` (see **SoC Tile Configuration** below)

## Key ESP Platform Conventions

### SoC Tile Configuration (`esp_defconfig`)

The ESP SoC is organized as a **NoC (Network-on-Chip) grid** of tiles. Each tile can be a CPU, memory, I/O, or accelerator. The tile layout is defined in `esp_xilinx-vc707-xc7vx485t_defconfig`.

The default configuration is a **2×2 grid**:
```
CONFIG_NOC_ROWS = 2
CONFIG_NOC_COLS = 2

TILE_0_0 = 0 mem mem
TILE_0_1 = 1 cpu cpu
TILE_1_0 = 2 empty empty
TILE_1_1 = 3 misc IO
```

To register an accelerator, replace an `empty` tile with an `acc` entry:
```
TILE_1_0 = 2 acc MYACCEL_SYSC_CATAPULT dma64 0 0 sld
```
Format: `TILE_<row>_<col> = <index> acc <ACCEL_NAME_UPPERCASE> dma64 0 0 sld`

If you need **more than one accelerator** and run out of empty tiles, you must **expand the NoC grid** by increasing `CONFIG_NOC_ROWS` and/or `CONFIG_NOC_COLS`, then add new tile entries for the additional slots. For example, a 2×3 grid:
```
CONFIG_NOC_COLS = 3

TILE_0_0 = 0 mem mem
TILE_0_1 = 1 cpu cpu
TILE_0_2 = 2 acc MY_ACCEL_1 dma64 0 0 sld
TILE_1_0 = 3 acc MY_ACCEL_2 dma64 0 0 sld
TILE_1_1 = 4 misc IO
TILE_1_2 = 5 empty empty
```

> **Note**: Expanding the NoC grid increases resource usage. Keep the grid as small as possible while fitting all required tiles.

### Other Conventions

- **Device IDs**: Each accelerator needs a unique ID in its XML file (e.g., `device_id="200"`). The same ID must match in the software driver's `_cfg.h` (as `0x200`). See **Device ID Allocation** below.
- **Register Offsets**: XML `<param>`s are mapped to MMIO register offsets **in reverse order**, starting at `0x40`. For example, if the XML lists three params `[P0, P1, P2]` (in that textual order), then `P2` sits at `0x40`, `P1` at `0x44`, and `P0` at `0x48`. The first-listed param always occupies the highest offset.
- **DMA Layout**: Data is laid out in `accel_shared_mem` as contiguous regions: `[input_A][input_B][output_O]`. Offsets must be aligned to `DMA_WORD_PER_BEAT`.
- **Fixed-point**: The accelerator uses fixed-point arithmetic. `float2fixed(x, WL, IL)` converts float to fixed-point with `WL` total bits and `IL` integer bits (fractional bits = WL - IL). Current config: WL=32, IL=16.
- **BSS is NOT zeroed**: On Ariane baremetal, static variables may contain garbage. Always explicitly initialize. `accel_common_init()` handles this for the shared infra.

### Device ID Allocation

- **Format**: The XML `device_id` attribute is a **hexadecimal** value written **without** the `0x` prefix. `device_id="200"` in XML corresponds to `0x200` in C code. **It is NOT decimal.** For example `device_id="04a"` ↔ `0x04a` in the driver.
- **Valid range**: `0x000` – `0x3FF` (10-bit field; see `devid_t` in the generated `sld_devices.vhd`).
- **Reserved by ESP** (do NOT use): `0x000`–`0x0FF` is densely populated by ESP infrastructure (caches, L2/LLC, TILE_CSR, etc.) and stock accelerators. `0x100` is NVDLA. `0x144` is sinkhorn. `0x222` is svd.
- **This project's reservation**: **`0x200` – `0x27F`**. The reference `dummy` accelerator occupies `0x200`. Assign your new accelerators starting at `0x201`.
- **Collision symptom**: If two accelerators share a device ID, `probe()` may return the wrong device or the SoC build may fail. Keep IDs unique.

## Restrictions

1. You are **NOT allowed** to access (read or write) files **outside the workspace directory**.
2. `dummy_workspace/` is a **reference-only** structural template. You must **not use it as-is** - its kernel is an identity pass-through and cannot perform any useful computation. Create your own accelerator workspace(s) by copying it and redesigning the HLS source, testbench, driver, and configuration to match your design.
3. Each accelerator must have a **unique device ID** in its XML file (`<your_workspace>/<accel_name>/<accel_name>.xml`).

## Evaluation Criteria

1. **Execution time**: Minimize total workload execution time (measured in cycles by `rdcycle64()`).
2. **Resource constraints** (VC707 FPGA):
   - Logic Cells: 485,760
   - Block RAM (Kb): 37,080
   - DSP Slices: 2,800
3. **Correctness**: The accelerated pipeline must **pass** whatever correctness check `systest.c` performs (e.g., output-vs-gold comparison, top-1 match, MSE threshold, etc.). The exact check and its threshold are workload-specific - look at what `systest.c` reports at the end of a run.

## Deliverables

Your job is done when you have completed all of the following:

1. Created accelerator HLS source(s) (new workspace directories with complete `src/`, `inc/`, `tb/`, and XML)
2. Written corresponding software driver(s) (header files with init/exec/cleanup functions)
3. Integrated drivers into `accel_drivers.h` and modified the workload's core computation to use your accelerators
4. Updated `esp_xilinx-vc707-xc7vx485t_defconfig` with accelerator tile entries (expanding NoC grid if needed)
5. Enabled `USE_ACCELERATOR` in `systest.c`

\end{markdownprompt}

\section{Performance Measurement}
\label{app:cycle}

We measure the cycle count of the application by inserting the following function:
\begin{lstlisting}[style=CStyle, caption=Cycle measurement on RISC-V CPU]
static inline uint64_t rdcycle64(void) {
    uint64_t val;
    __asm__ __volatile__("csrr %0, mcycle" : "=r"(val));
    return val;
}
\end{lstlisting}

\section{Evaluation Scripts}
\label{app:scripts}
We provide several scripts for automated evaluation and testing on the FPGA. Detailed usage instructions and setup requirements are available in the repository's \texttt{README}. The core automated workflow consists of:

\begin{enumerate}
    \item \texttt{check\_workspace.sh} (and associated utilities): This validation script handles environment initialization, behavioral simulation, HLS, software compilation, and installation into the ESP system. Because the LLM agent is isolated in a Docker container without access to the global ESP state, it may generate overlapping accelerator names or device IDs. This flow automatically invokes \texttt{check\_device\_ids.py} to resolve conflicts and \texttt{install\_registry.py} to manage the ESP tech-tree manifest, preventing integration failures.

    \item \textbf{Result Extraction and Analysis}: To collect and evaluate experimental metrics, the framework automatically processes raw logs and Vivado reports to generate structured CSV matrices. This includes parsing real FPGA execution traces from the UART port for performance evaluation, extracting resource utilization from hardware synthesis reports, tracking LLM API usage costs, and computing the overall hardware acceleration speedup. Relevant scripts are put under \texttt{results/}

    \item \texttt{run\_benchmark.sh}: The primary driver script that iterates through combinations of LLM models and algorithms. It integrates the \texttt{mini-swe-agent} flow with the ESP hardware flow to rebuild workspaces, invoke the agent, validate code, and deploy to the FPGA. It supports execution flags for partial runs, such as \texttt{--dry-run}, \texttt{--skip-fpga}, \texttt{--build-only}, and specific \texttt{--models} / \texttt{--algos} filters.
\end{enumerate}

\section{API Cost}
\label{app:total-cost}
\begin{table}[htbp]
  \centering
  \caption{Experimental Cost Summary Across Different Models}
  \label{tab:total-cost}
  \begin{tabular}{lc} 
    \toprule
    \textbf{Model} & \textbf{Total Cost (USD)} \\
    \midrule
    MiniMax M2.5   &  1.83\\
    Kimi K2.5      &  6.75\\
    Opus 4.6       &  46.96\\
    GPT-5.4        &  11.61\\
    Gemini 3.1 Pro &  73.88\\
    \bottomrule
  \end{tabular}
\end{table}

\section{BERT-Tiny Opus}
\label{app:bert_opus}

\begin{lstlisting}[style=CStyle, caption=Software-managed tiling logic generated by Claude Opus 4.6]
/* Initialize output with bias (or zero) */
for (int b = 0; b < batch_size; b++) {
    for (int i = 0; i < out_features; i++) {
        output[b * out_features + i] = (bias != 0) ? bias[i] : 0.0f;
    }
}

/* Tile over all three dimensions */
for (int c_start = 0; c_start < out_features; c_start += T) {
    int c_tile = out_features - c_start;
    if (c_tile > T) c_tile = T;

    for (int k_start = 0; k_start < in_features; k_start += T) {
        int k_tile = in_features - k_start;
        if (k_tile > T) k_tile = T;

        for (int r_start = 0; r_start < batch_size; r_start += T) {
            int r_tile = batch_size - r_start;
            if (r_tile > T) r_tile = T;

            /* Execute one tile: bert_tiny_vc707_opus_matmul_tile_buf = input_tile @ weight_tile^T */
            bert_tiny_vc707_opus_matmul_accel_tile(
                input + r_start * in_features + k_start, in_features,
                weight + c_start * in_features + k_start, in_features,
                bert_tiny_vc707_opus_matmul_tile_buf, c_tile,
                r_tile, c_tile, k_tile
            );

            /* Accumulate into output */
            for (int i = 0; i < r_tile; i++) {
                for (int j = 0; j < c_tile; j++) {
                    output[(r_start + i) * out_features + (c_start + j)] += bert_tiny_vc707_opus_matmul_tile_buf[i * c_tile + j];
                }
            }
        }
    }
}
\end{lstlisting}

\section{Autoencoder Opus}
\label{app:autoencoder_opus}

\begin{lstlisting}[style=CStyle, caption=Driver part of dense\_bn\_relu]
// Fuse batchnorm into weights and bias, convert to fixed-point
for (int i = 0; i < out_features; i++) {
    float scale = bn_gamma[i] / autoencoder_vc707_opus_dbnr_sqrtf(bn_var[i] + eps);
    
    // Fused weights: scale * weight
    for (int j = 0; j < in_features; j++) {
        float eff_w = scale * weight[i * in_features + j];
        accel_shared_mem[A_offset + i * in_features + j] = float2fixed(eff_w, AUTOENCODER_VC707_OPUS_DBNR_FX_WL, AUTOENCODER_VC707_OPUS_DBNR_FX_IL);
    }
    
    // Fused bias: scale * (bias - mean) + beta
    float eff_b = scale * (bias[i] - bn_mean[i]) + bn_beta[i];
    accel_shared_mem[B_offset + in_features + i] = float2fixed(eff_b, AUTOENCODER_VC707_OPUS_DBNR_FX_WL, AUTOENCODER_VC707_OPUS_DBNR_FX_IL);
}

// Load input vector
for (int j = 0; j < in_features; j++) {
    accel_shared_mem[B_offset + j] = float2fixed(input[j], AUTOENCODER_VC707_OPUS_DBNR_FX_WL, AUTOENCODER_VC707_OPUS_DBNR_FX_IL);
}

#undef AUTOENCODER_VC707_OPUS_DBNR_ROUND_UP_BEAT2

// Run accelerator
autoencoder_vc707_opus_dbnr_run_accel(in_features, out_features, apply_relu);

// Read output - convert from fixed-point
for (int i = 0; i < out_features; i++) {
    output[i] = fixed2float(accel_shared_mem[O_offset + i], AUTOENCODER_VC707_OPUS_DBNR_FX_WL, AUTOENCODER_VC707_OPUS_DBNR_FX_IL);
}
\end{lstlisting}

\begin{lstlisting}[style=CStyle, caption=Accelerator implementation (SystemC code)]
for(uint32_t i = 0; i < out_features; i++){
    // Start with bias value
    ACC_FPDATA acc = B_ping[0][in_features + i];
    
    // Dot product: weight_row_i . input
    #pragma hls_pipeline_init_interval 1
    for(uint32_t j = 0; j < in_features; j++){
        acc += A_ping[0][i * in_features + j] * B_ping[0][j];
    }
    
    // Convert accumulator back to output precision
    FPDATA result = acc;
    
    // Optional ReLU
    if(do_relu && result < 0){
        result = 0;
    }
    
    O_ping[0][i] = result;
}
\end{lstlisting}

\section{NightVision Opus}
\label{app:nightvision_opus}

\begin{lstlisting}[style=CStyle, caption=Accelerator implementation (SystemC code)]
for(uint32_t r = 1; r < rows - 1; r++){
    for(uint32_t c = 1; c < cols - 1; c++){
        // Read 3x3 window
        FPDATA w0 = A_ping[0][(r-1)*cols + (c-1)];
        FPDATA w1 = A_ping[0][(r-1)*cols + (c)];
        FPDATA w2 = A_ping[0][(r-1)*cols + (c+1)];
        FPDATA w3 = A_ping[0][(r)*cols   + (c-1)];
        FPDATA w4 = A_ping[0][(r)*cols   + (c)];
        FPDATA w5 = A_ping[0][(r)*cols   + (c+1)];
        FPDATA w6 = A_ping[0][(r+1)*cols + (c-1)];
        FPDATA w7 = A_ping[0][(r+1)*cols + (c)];
        FPDATA w8 = A_ping[0][(r+1)*cols + (c+1)];

        // Compare-and-swap macro
        #define CAS(a, b) do { \
            if ((a) > (b)) { FPDATA _t = (a); (a) = (b); (b) = _t; } \
        } while(0)

        // Bubble sort: 8 passes, fully sorts 9 elements
        // Pass 1 (8 CAS)
        CAS(w0, w1); CAS(w1, w2); CAS(w2, w3); CAS(w3, w4);
        CAS(w4, w5); CAS(w5, w6); CAS(w6, w7); CAS(w7, w8);
        // Pass 2 (7 CAS)
        CAS(w0, w1); CAS(w1, w2); CAS(w2, w3); CAS(w3, w4);
        CAS(w4, w5); CAS(w5, w6); CAS(w6, w7);
        // Pass 3 (6 CAS)
        CAS(w0, w1); CAS(w1, w2); CAS(w2, w3); CAS(w3, w4);
        CAS(w4, w5); CAS(w5, w6);
        // Pass 4 (5 CAS)
        CAS(w0, w1); CAS(w1, w2); CAS(w2, w3); CAS(w3, w4);
        CAS(w4, w5);
        // Pass 5 (4 CAS) - enough to get median at position 4
        CAS(w0, w1); CAS(w1, w2); CAS(w2, w3); CAS(w3, w4);
        // We only need w4 to be the median (5th smallest).
        // After 5 passes of bubble sort, the top 5 elements
        // (w4..w8) are in their correct positions.
        // Actually, bubble sort pushes the largest to the end.
        // After pass k, the k-th largest is in position (8-k+1).
        // After 5 passes: positions 4,5,6,7,8 are correct.
        // Wait, let me reconsider...
        // Actually, after each pass of bubble sort, the largest
        // unsorted element bubbles to its correct position.
        // After pass 1: w8 is correct (largest)
        // After pass 2: w7 is correct
        // After pass 3: w6 is correct
        // After pass 4: w5 is correct
        // After pass 5: w4 is correct (median!)
        // So we only need 5 passes instead of 8!

        #undef CAS

        // w4 is now the median (5th smallest)
        O_ping[0][r * cols + c] = w4;
    }
}
\end{lstlisting}

\end{document}